\documentclass[10pt,twocolumn,prl,aps,amssymb,amsmath,superscriptaddress]{revtex4}
\include{gyrokinetics-macros}
\usepackage{color,graphicx,ulem,bm}

\newcommand{\mbf}[1]{\mathbf{#1}}

\newcommand{\unit}[1]{\mathbf{\hat{#1}}}

\newcommand{\vth}[1][s]{\ensuremath{v_{\mathrm{th}_#1}}}

\newcommand{\gyroR}[2][s]{\ensuremath{{\left< #2 \right>}_\mathbf{R_\mathrm{#1}}}}
\renewcommand{\eqref}[1]{Eq.\ (\ref{#1})}

\renewcommand{\vth}{\ensuremath{ v_{\mathrm{th}} } }
\renewcommand{\gyroR}[1]{\ensuremath{{\left< #1 \right>}_{\bm{R}}}}

\newcommand{\la}{\left|}
\newcommand{\ra}{\right|}

\newcommand{\vvol}{d^{3}\mbf{v}}

\newcommand{\vpa}{v_{\parallel}}
\newcommand{\vpe}{v_{\perp}}

\newcommand{\beq}{\begin{equation}}
\newcommand{\eeq}{\end{equation}}
\newcommand{\grad}{\nabla}

\newcommand{\tprim}{\kappa}
\newcommand{\gexbcrit}{\gamma_E^{(c)}}

\begin{document}

\title{Turbulent transport in tokamak plasmas with rotational shear}

\author{M.\ Barnes}
\email{michael.barnes@physics.ox.ac.uk}
\affiliation{Rudolf Peierls Centre for Theoretical Physics, University of Oxford, 1 Keble Road, Oxford OX1 3NP, UK}
\affiliation{Euratom/CCFE Fusion Association, Culham Science Centre, Abingdon OX14 3DB, UK}
\author{F.\ I.\ Parra}
\affiliation{Rudolf Peierls Centre for Theoretical Physics, University of Oxford, 1 Keble Road, Oxford OX1 3NP, UK}
\author{E.\ G.\ Highcock}
\affiliation{Rudolf Peierls Centre for Theoretical Physics, University of Oxford, 1 Keble Road, Oxford OX1 3NP, UK}
\affiliation{Euratom/CCFE Fusion Association, Culham Science Centre, Abingdon OX14 3DB, UK}
\author{A.\ A.\ Schekochihin}
\affiliation{Rudolf Peierls Centre for Theoretical Physics, University of Oxford, 1 Keble Road, Oxford OX1 3NP, UK}
\author{S.\ C.\ Cowley}
\affiliation{Euratom/CCFE Fusion Association, Culham Science Centre, Abingdon OX14 3DB, UK}
\author{C.\ M.\ Roach}
\affiliation{Euratom/CCFE Fusion Association, Culham Science Centre, Abingdon OX14 3DB, UK}

\begin{abstract}

Nonlinear gyrokinetic simulations have been conducted to investigate turbulent transport in tokamak plasmas with rotational shear.  
At sufficiently large flow shears, linear instabilities are suppressed, but transiently growing modes drive
subcritical turbulence whose amplitude increases with flow shear.  This leads to a local minimum in the heat flux, indicating an optimal $\mbf{E}\times \mbf{B}$
shear value for plasma confinement.  Local maxima in the momentum fluxes are also observed, allowing for the possibility of bifurcations in the $\mbf{E}\times \mbf{B}$ shear.  
The sensitive dependence of heat flux on temperature gradient is relaxed for large flow shear values, with the critical temperature gradient increasing at lower flow shear values.
The turbulent Prandtl number is found to be largely independent of temperature and flow gradients, with a value close to unity.

\end{abstract}



\maketitle

\paragraph{Introduction.}

Experimental measurements in magnetic confinement fusion devices indicate that sheared 
mean $\mbf{E}\times \mbf{B}$ flows can significantly reduce and sometimes fully suppress turbulent particle,
momentum, and heat fluxes~\cite{burrellPoP97,conwayPRL00}.  Since these turbulent fluxes
determine mean plasma density and temperature profiles, their reduction leads to a local increase in
the profile gradients.  This increase can be dramatic:  
transport barriers in both the plasma core and edge
have been measured with radial extents on the order of only tens of ion Larmor radii~\cite{tressetNF02}.
The associated increase in core density and temperature results in increased fusion power.  
Thus, understanding how shear flow layers develop and what effect they have on 
turbulent fluxes is both physically interesting and practically useful.

This Letter reports a numerical study of the influence of sheared toroidal rotation on turbulent heat and momentum transport in tokamak plasmas.
Two main effects of sheared toroidal rotation were identified in previous numerical work~\cite{waltzPoP98,kinseyPoP05,peetersPoP05,cassonPoP09,roachPPCF09}: 
suppression of turbulent transport by shear in the perpendicular (to the mean magnetic field) velocity and linear destabilization due to the parallel
velocity gradient (PVG).  While the former observation indicates that a finite flow shear improves plasma confinement, the latter raises the question of
whether more shear is always beneficial.  Below we report that the PVG-driven linear instability~\cite{cattoPoF73} is stabilized at sufficiently
large flow shear values, consistent with fluid theory in slab geometry~\cite{newton}.  Correspondingly, fluxes decrease with increasing flow shear as the 
linear stabilization point is approached.
However, beyond this point, transiently
growing modes driven by the PVG give rise to subcritical turbulence.  The fluxes associated with this turbulence increase with flow shear.  
This implies an optimal flow shear
for each temperature gradient; the fact that the minimum heat flux value is finite indicates that there is a maximum attainable temperature gradient that can be maintained
for a given heat flux.  Additionally, the observed presence of maxima in the momentum fluxes
admits the possibility of bifurcations in the flow shear (and thus the temperature gradient).

In the absence of flow shear, a small increase in temperature gradient leads to a large increase in heat flux (``stiff transport").
Recent experimental evidence~\cite{manticaPRL09} suggests that flow shear may reduce
this sensitivity in configurations with low magnetic shear.  Our results indicate that at low flow shear values, both the critical temperature gradient
for the onset of turbulence and the stiffness increase.  At high flow shear values, the opposite behavior is observed (the stiffness and critical temperature gradient
both decrease).

\paragraph{Model.}

A closed set of evolution equations~\cite{sugamaPoP97,abelPPCF10} for mean plasma density, pressure, and toroidal angular momentum is obtained 
by taking moments of the kinetic equation and applying the $\delta f$ gyrokinetic ordering~\cite{antonPoF80,friemanPoF82}.  
These moment equations relate the evolution of the mean quantities to particle, momentum, and heat fluxes, which are typically dominated by turbulent contributions.  
We restrict our attention to electrostatic fluctuations and assume a modified Boltzmann response~\cite{hammettPPCF93} for the electron distribution.  
The resultant particle flux is 
identically zero, and the radial components of the turbulent heat flux of species $s$, $Q_s$, and toroidal angular momentum flux, $\Pi$, are given by
\begin{eqnarray}
Q_s&=&\overline{\int\vvol \left(\frac{m_s v^2}{2}\right) \left(\mbf{v}_{E}\cdot\frac{\grad\psi}{\overline{\la\grad\psi\ra}}\right)\delta f_s} \label{eqn:qflx}\\
\Pi&=&\overline{\sum_s m_sR^2\int\vvol \left(\mbf{v}\cdot\grad{\phi}\right) \left(\mbf{v}_{E}\cdot\frac{\grad\psi}{\overline{\la\grad\psi\ra}}\right)\delta f_s}, \label{eqn:vflx}
\end{eqnarray}
where $\psi$ is a flux-surface label, $\phi$
the toroidal angle, $m_s$ the particle mass, $\mbf{v}$ its velocity in the frame of mean flow, $R$ its major radius,
$\mbf{v}_E$ the fluctuating $\mbf{E}\times \mbf{B}$ drift velocity, $\delta f_s$ the deviation of the 
distribution function from a local Maxwellian, and the overline denotes a spatial average over a thin annular region encompassing a given flux surface.

The distribution function $\delta f_s$ appearing in Eqs.~(\ref{eqn:qflx}) and~(\ref{eqn:vflx}) is calculated 
by solving the standard $\delta f$ gyrokinetic equation in the limit where the plasma flow speed, $u$, is ordered comparable to the 
ion thermal speed, $\vth$~\cite{artunPoP94,abelPPCF10}.  In this ``high-flow" regime,
the flow velocity is constrained to be $\mbf{u}=R^2\omega(\psi)\grad\phi$, where $R$ is the major radius and 
$\omega$ is the rotational frequency~\cite{hintonPoF82}.  We consider a subsidiary expansion 
in low Mach number $M$ ($\rho_i/L \ll M \ll 1$), where $L$ is a typical macroscopic length scale (e.g., the minor radius of the torus) and $\rho_i$ is the ion Larmor radius.
However, we allow for flow gradients of order $1/ML$, and so neglect terms proportional 
to $\omega$ (such as Coriolis and centrifugal drifts) but retain those proportional to $d\omega/d\psi$ (flow shear).  The 
resulting equation is
\beq
\begin{split}
\label{eqn:gk}
&\frac{dh}{dt} + \left(\vpa\unit{b} + \mbf{v}_B+ \gyroR{\mbf{v}_E }\right) \cdot\grad h - \gyroR{ C[h] }= \\
& - \gyroR{ \mbf{v}_E } \cdot\grad\psi\left(\frac{d F_0 }{d\psi} + \frac{m\vpa}{T} \frac{RB_{\phi}}{B}\frac{d\omega}{d\psi}F_0\right) + \frac{eF_0}{T}\frac{d\gyroR{\varphi}}{dt},
\end{split}
\eeq
where $d/dt=\partial/\partial t +\mbf{u}\cdot\grad$, $h=\delta f + (e\varphi/T) F_0$ is the deviation of $\delta f$ from a Boltzmann response, $e$ is particle charge, $\varphi$ is the electrostatic potential, $\mbf{v}_E=(c/B^2)\mbf{B}\times\grad\varphi$, $T$ is temperature, $F_0$ is a Maxwellian distribution of peculiar velocities, $C$ is the collision operator, $B$ is the magnetic field strength (with $B_{\phi}$ the toroidal component), $\mbf{v}_B=\unit{b}/\Omega \times \left(\vpa^2\unit{b}\cdot\grad\unit{b}+\vpe^2 \grad{B}/2B\right)$ contains magnetic drifts, $\Omega$ 
is the Larmor frequency, $\unit{b}=\mbf{B}/B$, and $\gyroR{.}$ denotes the average over gyro-angle at fixed guiding center
position $\mbf{R}$.

\begin{figure}
\includegraphics[height=2.2in]{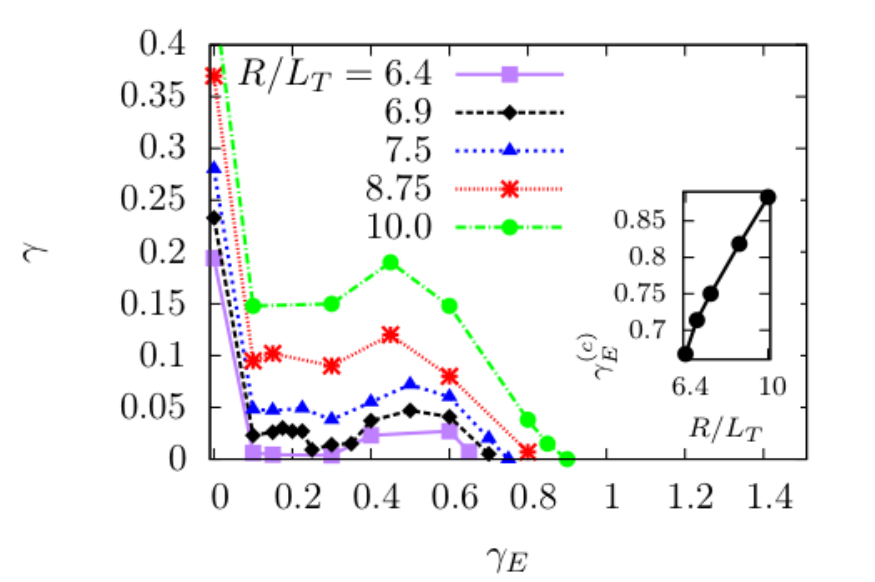}
\caption{Average linear growth rates vs. flow shear for different values of $R_0/L_T$.  Inset: critical flow shear vs. $R_0/L_T$.}
\label{fig:gamvsgexb}
\end{figure}

Eq.~(\ref{eqn:gk}) is solved in the rotating reference frame using the local, nonlinear gyrokinetic code \texttt{GS2}~\cite{kotschCPC95,roachPPCF09}.
The toroidal velocity is expanded in $\psi$ about its value at the center of the simulation domain, $\psi_0$, giving $\mbf{u}\approx R^2(\psi-\psi_0)(d\omega/d\psi)\grad\phi$.
The local approximation assumes that $d\omega/d\psi$ is constant across the simulation domain.  Consequently, the effect of sheared flow is accounted for by a single parameter, $\gamma_E=(\psi/q)(d\omega/d\psi)R_0/v_{th}=(M/q)(d\ln\omega/d\ln r)$, where $q$ is the safety factor, $R_0$ is the major radius at the 
center of the flux surface, and $r$ is the half-diameter of the flux surface (both measured at the height of the magnetic axis).

We study a system whose magnetic geometry corresponds to the widely-used Cyclone base case~\cite{dimitsPoP00} 
(unshifted, circular flux surface with $q=1.4$, magnetic shear $\hat{s}=d\ln q/d\ln r=0.8$, $r/R_0=0.18$, and $R_0/L_n$ = 2.2, with
$L_n^{-1}=-d\ln n/dr$).
The $\mbf{E}\times \mbf{B}$ shearing rate, $\gamma_E$, and the normalized inverse temperature gradient scale length, $\tprim\equiv R_0/L_T$,
were varied over a wide range of values in a series of linear and nonlinear simulations.

\paragraph{Linear stability.} The average linear growth rates, $\gamma$, obtained from these simulations are given in Fig.~\ref{fig:gamvsgexb}.  
We see that $\gamma$ decreases rapidly with $\gamma_E$ for small values of $\gamma_E$ before increasing to a local maximum and subsequently decreasing to zero.  
For $\gamma_E\gtrsim0.25$, the
system is linearly unstable only in the presence of both temperature and parallel flow gradients (cf. [\onlinecite{kinseyPoP05,peetersPoP05,cassonPoP09,roachPPCF09}]).
The system becomes linearly stable for large values of $\gamma_E$, with the critical $\gamma_E$ for stability, $\gexbcrit$,
increasing approximately linearly with $\tprim$.  Beyond $\gexbcrit$, there are no linearly unstable modes.  This result is qualitatively
similar to the prediction from fluid theory in a slab~\cite{newton}, where no linear instability is possible when $\gamma_E/\hat{s}$ 
exceeds a certain critical value.  

\begin{figure}
\includegraphics[height=2.2in]{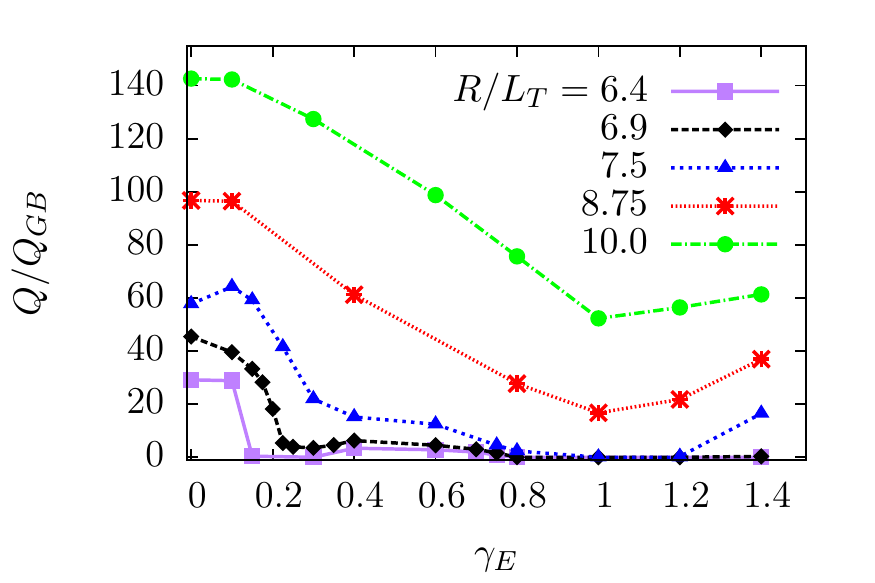}
\includegraphics[height=2.2in]{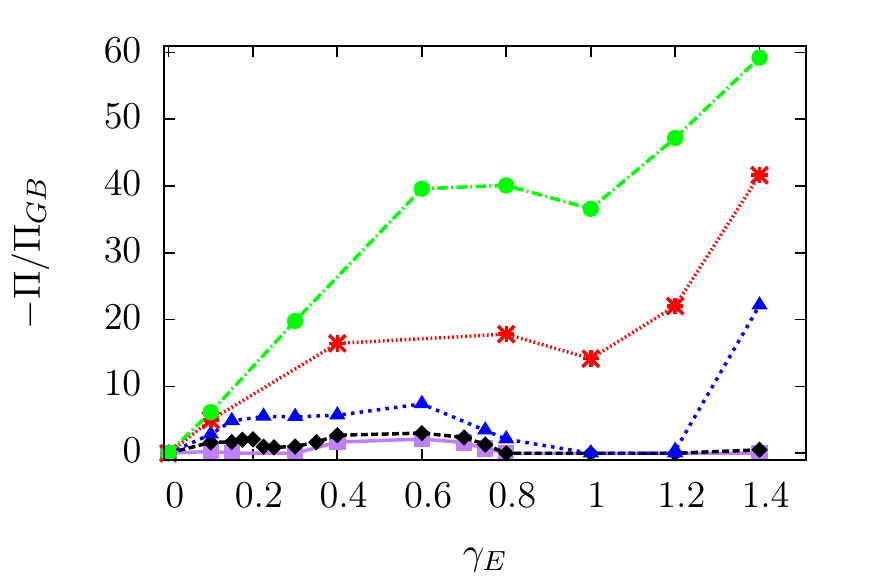}
\caption{Turbulent heat (top) and toroidal angular momentum (bottom) fluxes vs. flow shear for various values of 
$R_0/L_T$.  For $\gamma_E\lesssim0.25$, the heat flux is
reduced by $E\times B$ shear.  Beyond this value, the PVG drives instability and both heat and momentum fluxes 
increase.  For $\gamma_E> \gexbcrit$, the turbulence is subcritical, sustained by transiently growing modes.}
\label{fig:flxvsgexb}
\end{figure}

\paragraph{Heat flux.} The turbulent heat flux calculated from nonlinear simulations is given in Fig.~\ref{fig:flxvsgexb}.  
For all $\tprim$ values, the heat flux qualitatively follows the same trend as growth rates when $\gamma_E < \gexbcrit$.  
For $\gamma_E>\gexbcrit$, nonlinear simulations initialized with low-amplitude noise develop no turbulent transport.  
However, for finite initial fluctuation amplitudes, one finds
that the turbulence does not necessarily decay away:  for sufficiently large values of $\tprim$ and initial amplitude, the flux reaches steady state
values in excess of those found for $\gamma_E$ just below $\gexbcrit$~\cite{kinseyfootnote}.  The flux then increases monotonically with 
$\gamma_E$.  For the range of $\gamma_E$ considered here, the heat flux associated with a given temperature gradient is thus minimized at a finite shearing rate.

\begin{figure}
\includegraphics[height=2.2in]{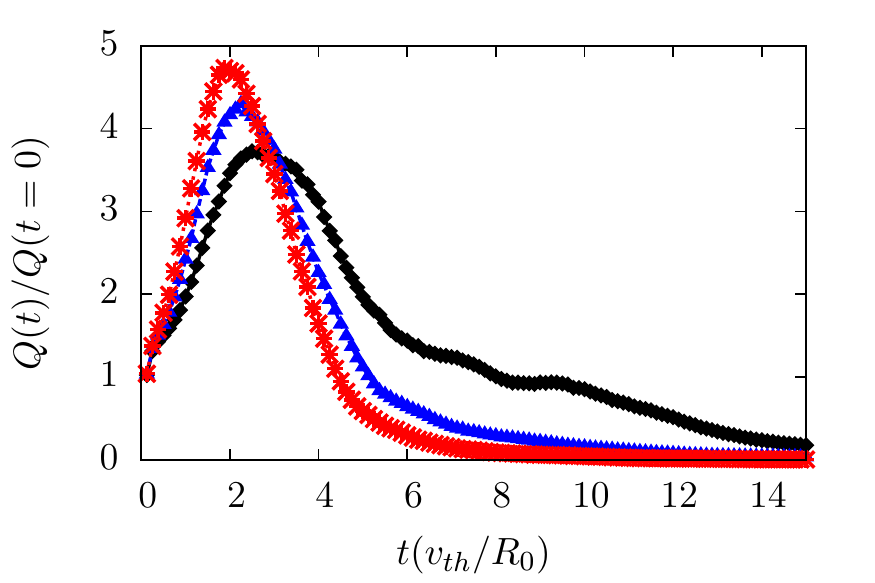}
\caption{Heat flux versus time obtained from linear simulations for $R_0/L_T=8.75$ and three $\gamma_E$ values at which subcritical turbulence is observed.}
\label{fig:transient}
\end{figure}

\paragraph{Subcritical turbulence.}  Because the turbulence is present in the absence of linear instability, we refer to it as subcritical.  
When $\gamma_E>\gexbcrit$, linear simulations exhibit transient growth, with
order unity increases in the initial fluctuation amplitudes over times of several $R_0/\vth$ before subsequent decay (Fig.~\ref{fig:transient}).  The duration of growth decreases
with increasing flow shear, but the transient growth rate increases so that the amplification factor of the initial perturbation amplitude grows with flow shear (cf.~\cite{newton}).
This transient amplification provides an energy source for the turbulence, which can be maintained by the nonlinearity through redistribution of energy
amongst other modes.  We have identified the PVG term in Eq.~\ref{eqn:gk} (the $d\omega/d\psi$ term on the RHS) as the driver of the subcritical turbulence, as the latter is no longer present 
when the PVG is artificially set to zero.

\paragraph{Momentum flux.} The toroidal angular momentum flux also mimics the linear growth rate, except near $\gamma_E=0$~\cite{peetersPRL07}.  
This is understood by expressing the momentum flux in diffusive form: $\Pi=-m_i\vth(qR_0/r)\nu_i(\kappa,\gamma_E)\gamma_E$, where $\nu_i$ is the turbulent viscosity.  For $\gamma_E$ small, the fluctuation amplitudes vary little, so $\nu_i$ is approximately constant and $\Pi\propto\gamma_E$.  For larger $\gamma_E$, turbulent amplitudes drop rapidly so that $\Pi$ decreases, resulting in the local maxima seen in Fig.~\ref{fig:flxvsgexb}.  
This suppression of $\Pi$ is due to linear stabilization as $\gamma_E$ approaches $\gexbcrit$ from below.
The momentum flux then increases monotonically when $\gamma_E>\gexbcrit$ due to subcritical turbulence.
The maxima in $\Pi$ may lead to a bifurcation in flow shear, discussed in the Conclusions.

\begin{figure}
\includegraphics[height=2.2in]{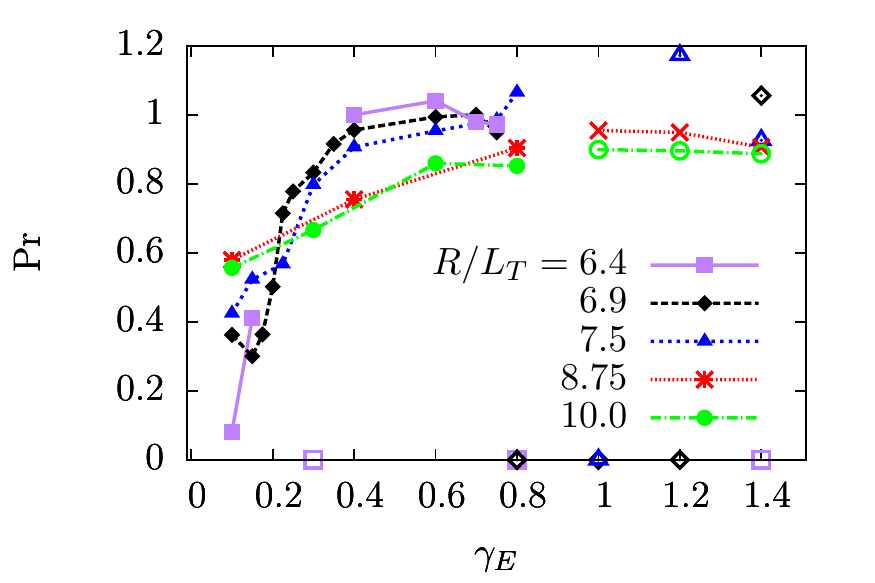}
\caption{Turbulent Prandtl number as a function of flow shear for various values of $R_0/L_T$.  Unfilled points correspond to linearly stable flow shear values.  Zero Pr points
represent fully suppressed turbulence.}
\label{fig:prandtl}
\end{figure}

\paragraph{Prandtl number.}  The turbulent Prandtl number can be calculated from the values of $Q_i$ and $\Pi$.  It is defined as 
$\textnormal{Pr}=\nu_i/\chi_i$, 
where the turbulent thermal diffusivity, $\chi_i$, is given by $Q_i=-\chi_i dT_i/dr$.  
Fig.~\ref{fig:prandtl} shows that $\textnormal{Pr}$ is approximately independent of $\tprim$ and only has strong dependence
on $\gamma_E$ for small values of $\gamma_E$~\cite{cassonfootnote}.  This is despite the fact that both $\nu_i$ and $\chi_i$ individually have strong dependencies on $\tprim$ and $\gamma_E$.  
For $\gamma_E\gtrsim0.4$ the Prandtl number is close to unity, in good agreement with experimental measurements at low Mach numbers~\cite{devriesPPCF10}.

\begin{figure}
\includegraphics[height=2.2in]{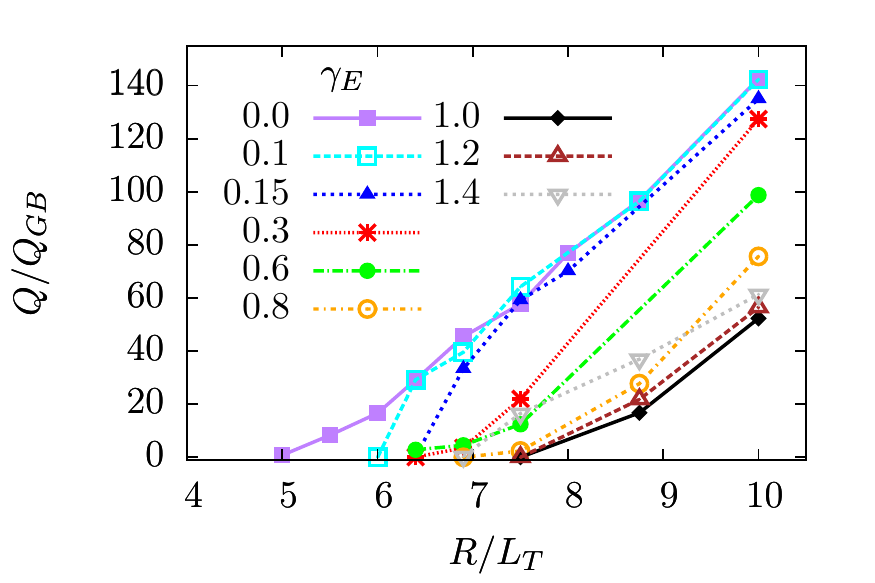}
\caption{Turbulent heat flux vs $R_0/L_T$ for various values of the flow shear.  Though there is some minimal relaxation of
the profile stiffness for $0.3\lesssim\gamma_E\lesssim0.8$, the predominant effect of the shear is to shift the critical temperature gradient.}
\label{fig:stiff}
\end{figure}

\paragraph{Stiff transport.} A serious impediment to confinement is the sensitive dependence of the heat flux on small changes in $\tprim$.  This
stiffness of the transport makes it difficult to increase $\tprim$, and therefore the core temperature, beyond the critical value, $\tprim_c$, at which turbulence is excited.  Recent
experimental results indicate that stiffness, $dQ_i/d\tprim$, may be reduced at low magnetic shear, $\hat{s}$, and large values of $\gamma_E$~\cite{manticaPRL09}.
For the Cyclone base case considered here ($\hat{s}=0.8$), we find a complicated dependence of stiffness on flow shear, as illustrated in Fig.~\ref{fig:stiff}.
At low values of $\gamma_E$ ($\lesssim 0.3$), the critical $\tprim$ shifts to higher values, but the stiffness increases.  This makes sense intuitively:  one expects that for 
$\tprim/\gamma_E\rightarrow\infty$, the heat flux will tend to the curve corresponding to $\gamma_E=0$.  For this to happen, the value of $dQ_i/d\tprim$ at $Q_i=0$ must increase to compensate for the increase in $\tprim_{c}$. 

For $0.3\lesssim\gamma_E < \gexbcrit$, both $\tprim_c$ and the profile stiffness decrease.  This is a result of
a change in the nature of the linear instability, which is now driven by the PVG.  The relaxation of stiffness is modest,
and it only occurs for $\tprim$ near $\tprim_c$.  When $\gamma_E>\gexbcrit$,
$\tprim_c$ initially shifts upwards with increasing $\gamma_E$, and stiffness increases for $\tprim$ near $\tprim_c$.  
However, for even larger $\gamma_E$, both the stiffness and $\tprim_c$ decrease.  This
is to be expected, as the subcritical turbulence is driven by the velocity, not temperature, gradient and thus has a weaker dependence on $\tprim$.

\paragraph{Conclusions.}

We have shown that PVG-driven turbulence exists at large flow shears, but only if the initial fluctuation amplitudes are sufficiently large.  This has a number
of potentially important implications.  First, it indicates that linear stability analysis in insufficient to determine critical gradients in rotating plasmas.
Furthermore, it shows that the system can undergo hysteresis: The fluxes associated with a given flow shear and temperature gradient pair
depend on the path taken to obtain that pair.  As an example, consider the pair $\gamma_E=1$, $R_0/L_T=10$.  From Fig.~\ref{fig:flxvsgexb}, we see
that if $R_0/L_T=10$ is obtained at $\gamma_E<\gamma_E^{(c)}$ before increasing $\gamma_E$ to unity, there will be large heat and momentum fluxes.  In this
case, there is an optimal flow shear for confinement ($\gamma_E\approx 1$) and a maximum attainable temperature gradient for a given power input.
However, if $\gamma_E=1$ is obtained at $R_0/L_T\lesssim7.5$ before increasing $R_0/L_T$ to ten, then the fluxes will be zero because the initial amplitude is not large enough
to excite the subcritical turbulence.  Thus it may be beneficial to
increase flow shear on a time scale short compared to the energy confinement time.


The existence of local maxima of the toroidal angular momentum flux provides the potential for bifurcations in $\gamma_E$ and corresponding
bifurcations in temperature gradient.  However, a detailed transport analysis is necessary to definitively observe such a bifurcation
for experimentally relevant conditions.    Finally, we note that the $\mbf{E}\times\mbf{B}$ suppression and PVG drive terms differ by a factor proportional
to $(qR_0/r)(B_{\phi}/B)$~\cite{roachPPCF09}.  Thus our results, which should be qualitatively robust, may undergo considerable quantitative
variation with changes in magnetic configuration.


The authors are grateful to I. G. Abel, F. J. Casson, W. Dorland, G. W. Hammett, and A. Zocco for useful discussions.  
M.B. was supported by the Oxford-Culham Fusion Research Fellowship.  The authors also thank the Leverhulme Trust (UK) 
International Network for Magnetized Plasma Turbulence for travel support.  Computing time for this research was provided
by EPSRC grant EP/H002081/1.


\begin{thebibliography}{22}
\bibitem[\citenamefont{Kinsey et~al.}(2009)\citenamefont{Kinsey et~al.}]{kinseyfootnote}
\bibinfo{note}{We have verified that this subcritical turbulence persists when kinetic electrons are included, in contrast with the qualitatively different phenomenon observed In [\onlinecite{kinseyPoP05}].}
\bibitem[\citenamefont{Casson et~al.}(2009)\citenamefont{Casson et~al.}]{cassonfootnote}
\bibinfo{note}{In [\onlinecite{cassonPoP09}], which considers a different magnetic equilibrium and flow shear values for which there is no subcritical turbulence, Pr is found to be unity and independent of $\gamma_E$ for a single $\tprim$ value.}
\end{thebibliography}

\end{document}